\RequirePackage{arydshln}
\documentclass[aps,twocolumn,nofootinbib,superscriptaddress,preprintnumbers,10pt,floatfix]{revtex4-1}

\usepackage[utf8]{inputenc}

\usepackage{float}
\usepackage{amsmath,amssymb}
\usepackage{dsfont} 
\usepackage{hyperref}
\usepackage{graphicx}
\usepackage{enumitem}
\usepackage{mathtools}
\usepackage{bbold}
\usepackage{multirow}
\usepackage{ytableau}
\usepackage{youngtab}
\usepackage{braket}
\usepackage{soul}
\usepackage{cancel}
\usepackage{xcolor}
\usepackage[normalem]{ulem}

\usepackage{lipsum}

\usepackage{colortbl} 
\definecolor{Gray}{gray}{0.95}
\definecolor{RGray}{gray}{0.90}
\definecolor{CGray}{gray}{0.92}
\definecolor{nicegreen}{rgb}{0.1,0.5,0.1}

\usepackage{arydshln}

\newcommand{\mrm}[1]{\mathrm{#1}}
\newcommand{\re}[0]{\mrm{Re}}

\makeatletter
\g@addto@macro\bfseries{\boldmath}
\makeatother

\makeatletter
\renewcommand\paragraph{\@startsection{paragraph}{4}{\z@}%
                                    {3.25ex \@plus1ex \@minus.2ex}%
                                    {-1em}%
                                    {\normalfont\normalsize\bfseries}}
\makeatother

\allowdisplaybreaks
\begin{document}

\preprint{}
\preprint{}

\title{ Right-handed interactions in puzzling $B$-decays}

\author{Damir Be\v{c}irevi\'c}
\email{damir.becirevic@ijclab.in2p3.fr}
\affiliation{IJCLab, P\^ole Th\'eorie (Bat.~210), CNRS/IN2P3 et Universit\'e Paris-Saclay, 91405 Orsay, France}
\author{Svjetlana Fajfer} 
\email{svjetlana.fajfer@ijs.si}
\affiliation{Department of Physics, University of Ljubljana, Jadranska 19, 1000 Ljubljana, Slovenia}
\affiliation{Jo\v zef Stefan Institute, Jamova 39, 1000  Ljubljana, Slovenia}
\author{Nejc Košnik} 
\email{nejc.kosnik@ijs.si}
\affiliation{Department of Physics, University of Ljubljana, Jadranska 19, 1000 Ljubljana, Slovenia}
\affiliation{Jo\v zef Stefan Institute, Jamova 39, 1000  Ljubljana, Slovenia}
\author{Lovre Pavičić} 
\email{lovre.pavicic@ijs.si}
\affiliation{Jo\v zef Stefan Institute, Jamova 39, 1000 Ljubljana, Slovenia}

\begin{abstract}
\vspace{5mm}
We explain the difference between the measured decay widths of $b \to c \tau \nu$ processes and of $B\to K\nu\bar{\nu}$ and their values predicted in the Standard Model by introducing the right-handed interactions both to quarks and to leptons. At low energy scales, in addition to the Standard Model particles, we assume the presence of an additional neutral lepton (right-handed neutrino). We then show a specific realization of such a scenario in a model with a single scalar leptoquark ($S_1$), and discuss the corresponding phenomenology.
\vspace{3mm}
\end{abstract}

\maketitle

\allowdisplaybreaks

\section{Introduction}\label{sec:intro}
In order to describe a puzzling difference between the measured and the Standard Model (SM) value of $R_{D^{(\ast )}} = \mathcal{B}(B\to D^{(\ast )} \tau\nu)/\mathcal{B}(B\to D^{(\ast )} l \nu)$, with $l\in \{e,\mu\}$, one needs to resort to physics beyond the Standard Model (BSM). That need is even more compelling if the deviation is attributed solely to $b \to c \tau \nu$ as it will become clear below. In addition to that puzzle, the result of the recently measured $\mathcal{B}(B^\pm\to K^\pm \nu\bar\nu)$ also turned out to be considerably larger than predicted in the SM~\cite{Belle-II:2023esi}. While none of these discrepancies are at the $5\sigma$ level, they were unexpected and need to be further scrutinized. 
Seeking a BSM scenario which can simultaneously accommodate both of these deviations becomes quite a complicated task if one sticks to a minimal extension of the SM. In particular, in Refs.~\cite{Allwicher:2023xba,Bause:2023mfe} it was shown that in terms of low energy effective theory (LEFT), one has to introduce the right-handed (RH) couplings to quarks to simultaneously agree with $\mathcal{B}(B^\pm\to K^\pm \nu\bar\nu)^\mathrm{exp}$ and with the experimental bound on $\mathcal{B}(B\to K^\ast \nu\bar\nu)^\mathrm{exp}$ (also, in~\cite{Allwicher:2024ncl} the correlation between $b\to s \nu\bar\nu$ and $s\to d \nu\bar\nu$ transitions was explored). Alternatively, that can be achieved by introducing a neutral fermion (RH neutrino) with a peculiar mixing with the SM neutrinos~\cite{Felkl:2023ayn,Rosauro-Alcaraz:2024mvx}.   
In this letter, we show that this can be achieved by the effective operators, which couple via RH couplings both to quarks and leptons. In other words, we will assume the existence of an RH neutrino with a mass that can be probed through both $b \to c \tau \nu$ and $b \to s \nu \nu$ processes. Before entering that topic, let us remind the reader that the most recent experimental averages~\cite{HFLAV:2022esi}, 
\begin{equation}\label{RDexp}
R_{D}^\mathrm{exp} = 0.344(26), \quad R_{D^*}^\mathrm{exp} = 0.285(12),
\end{equation}
are respectively about $2\sigma$ and $3\sigma$ larger than their SM counterparts, $R_D^\mathrm{SM}=0.293(8)$~\cite{FlavourLatticeAveragingGroupFLAG:2021npn} and  $R_{D^*}^\mathrm{SM} = 0.247(2)$~\cite{Becirevic:2024pni}.
As for the $b \to s \nu \bar \nu$ processes, the recently measured $\mathcal{B}(B^\pm\to K^\pm \nu \bar\nu)^\mathrm{exp}=2.35(67)\times 10^{-5}$~\cite{Belle-II:2023esi} is  larger than predicted, $\mathcal{B}(B^\pm\to K^\pm \nu \bar\nu)^\mathrm{SM}=4.4(3)\times 10^{-6}$~\cite{Buras:2014fpa,Becirevic:2023aov}.
For shortness, in what follows we will use $R_{K^{(\ast )}}^{\nu\nu}= 
\mathcal{B}(B \to K^{(\ast )} \nu \bar\nu)/\mathcal{B}(B \to K^{(\ast )} \nu \bar\nu)^\mathrm{SM}$, so that we have~\cite{Belle-II:2023esi}
\begin{equation}\label{RKexp}
R_{K^+}^{\nu\nu\,\mathrm{(exp)}}=5.4\pm 1.5\,.
\end{equation}
Notice that the old Belle measurement~\cite{Belle:2017oht} indicates the upper bound of $R_K^{\text{inv}}<3.6$, slightly in tension with the result eq.~\eqref{RKexp}. In this letter we take as input only the above Belle II result.

\section{LEFT and Phenomenology}\label{sec:eff}

We now briefly remind the reader of the LEFT both in the SM and in a minimal BSM scenario that we are proposing here.

\subsection{SM LEFT}
In the SM  $b \to c \ell \nu$ transitions, $\ell\in \{e,\mu,\tau\}$, are the tree level processes described by
\begin{equation}
    \label{eq:Leffbc}
\mathcal{L}_\mathrm{eff}^\mathrm{b\to c\ell \nu} = 
- \dfrac{4 G_F}{\sqrt{2}} V_{cb} \left(\bar{c} \gamma^\mu P_L b\right)\left(\bar{\ell} \gamma_\mu P_L \nu \right) +\mathrm{h.c.}\,,
\end{equation}
where $P_{L,R}=(1\mp\gamma_5)/2$, $G_F$ is the Fermi constant, and $V_{cb}$ a relevant entry in the Cabibbo--Kobayashi-Maskawa (CKM) matrix. 
The main obstacle to the precision evaluation of $\mathcal{B}(B\to D^{(\ast )} \ell\nu)$ is related to the hadronisation effects described by nonperturbative QCD. While 
the corresponding hadronic matrix elements can be computed by using lattice QCD in the case of $D$-meson in the final state~\cite{FlavourLatticeAveragingGroupFLAG:2021npn}, this is much more complicated in a similar decay to $D^\ast$~\cite{LATTICES}. For that reason, and by assuming that $R_{D^*}^\mathrm{exp}>R_{D^*}^\mathrm{SM}$ is completely due to the enhancement of $\mathcal{B}(B\to D^{\ast } \tau\nu)$, we can use the form factors extracted from experimental analyses of the angular distribution of $B\to D^{\ast }( \to D\pi) l\nu$~\cite{HFLAV:2022esi}, with a minimal additional input from lattice QCD, as explained in Ref.~\cite{Becirevic:2024pni}.\\
In contrast, $b \to s \nu \nu$ transitions are loop-induced in the SM and at low energy scales, they are described by
\begin{equation}
\label{eq:eft-bsnunu}
\mathcal{L}_\mathrm{eff}^{\mathrm{b\to s\nu\nu}} =  - \dfrac{4 G_F}{\sqrt{2}} \frac{\alpha_\mathrm{em}\lambda_t}{ 2 \pi} C_L^\mathrm{SM} (\bar{s} \gamma_\mu P_L b)\left(\bar{\nu} \gamma^\mu P_L \nu \right) +\mathrm{h.c.}\,, 
\end{equation}
where $\alpha_\mathrm{em}=e^2/4\pi$, $\lambda_t = V_{tb}^\ast V_{ts}$, and $C_L^\mathrm{SM}=6.32(7)$~\cite{Brod:2010hi}. As for the hadronic matrix elements, those entering the $B\to K$ transitions are well described by lattice QCD~\cite{HFLAV:2022esi}, while for $K^\ast$ in the final state the situation is not as clear and the light cone QCD sum rule (LCSR) results are used with lattice QCD result at large $q^2$ used as a constraint~\cite{Bharucha:2015bzk}. 

\subsection{Our proposal}

In addition to the effective Lagrangians written above, we need a BSM contribution to accommodate the observations that  $r_{D^{(\ast )}} = R_{D^{(\ast )}}^\mathrm{exp}/R_{D^{(\ast )}}^\mathrm{SM}> 1$, $R_{K}^{ \text{`inv'}\,\mathrm{(exp)}}> 1$, and that $R_{K^{\ast }}^{ \text{`inv'}\,\mathrm{exp}}$ is bounded from above, cf. Eqs.(\ref{RDexp},\ref{RKexp}). This can be achieved by a simple extension of the SM by 
\begin{align}
\label{eq:RR}
\mathcal{L}_\mathrm{eff} \supset &\, \mathcal{L}^{b\to c \tau N_R} +  \mathcal{L}^{b\to s N_R N_R} \nonumber \\
=& - \sqrt 2 G_F \,C_{RR} ( \bar c \gamma_\mu P_R b ) (\bar \tau  \gamma^\mu P_R N_R)\\
& - \sqrt 2 G_F \widetilde C_{RR} ( \bar s \gamma_\mu P_R b ) (\bar N_R \gamma^\mu P_R N_R) +\mathrm{h.c.},\nonumber
\end{align}
where we assume the existence of an RH neutral lepton $N_R$ (RH neutrino) and of a heavy mediator that couples both to quarks and leptons via the RH currents at the low energy scales. $C_{RR}$ and $\widetilde C_{RR}$ are the two BSM couplings (Wilson coefficients) relevant to $b\to c\tau N_R$ and $b\to s N_R  N_R$ transitions, respectively. They are a priori independent from each other but can be related in a particular model compatible with the above Lagrangian.\footnote{Note that in the RH effective interactions it is natural not to use the SM normalization with $V_{cb}$.}

\begin{figure}[t!]
\centering
\includegraphics[width=.91\linewidth]{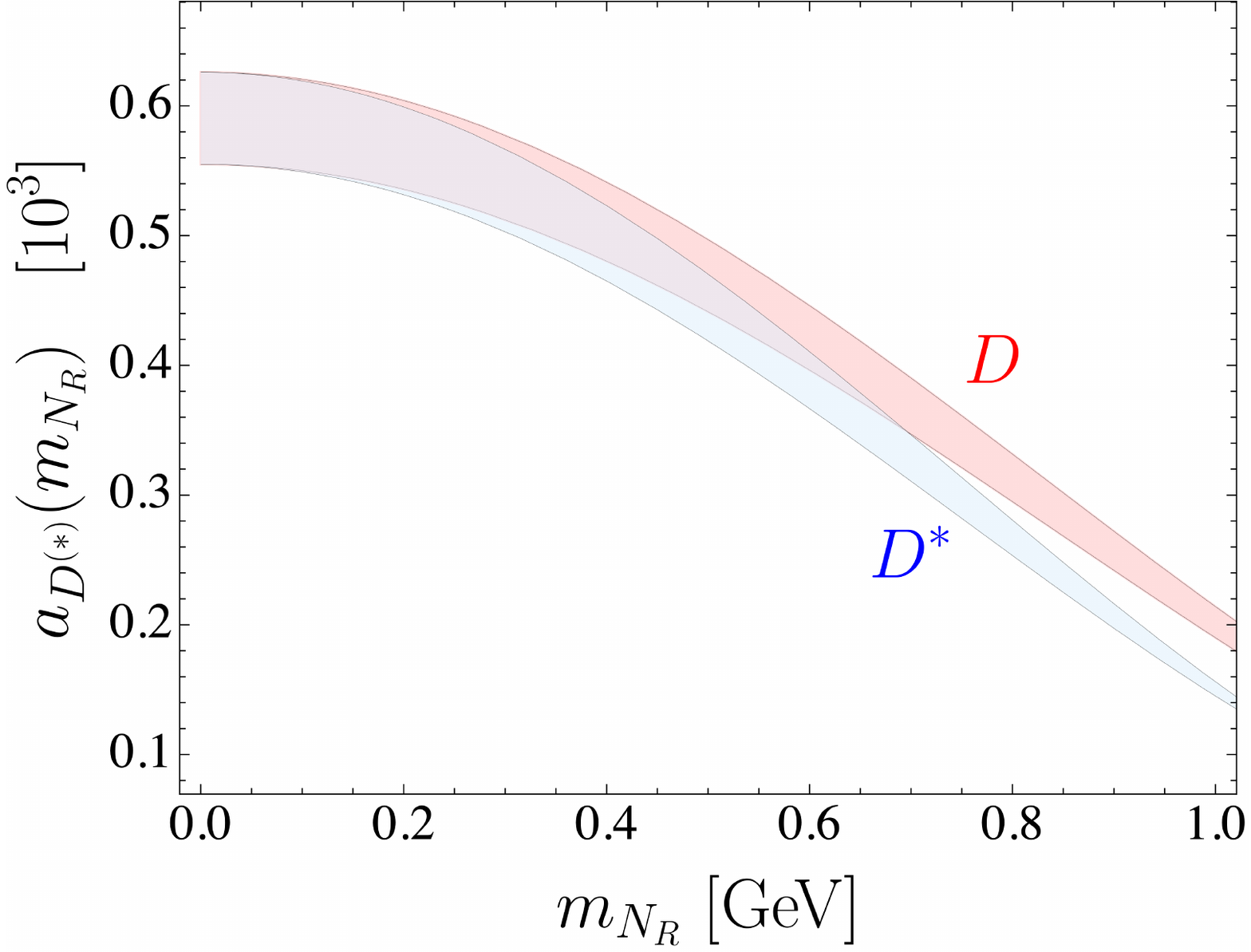}\\
\includegraphics[width=0.85\linewidth]{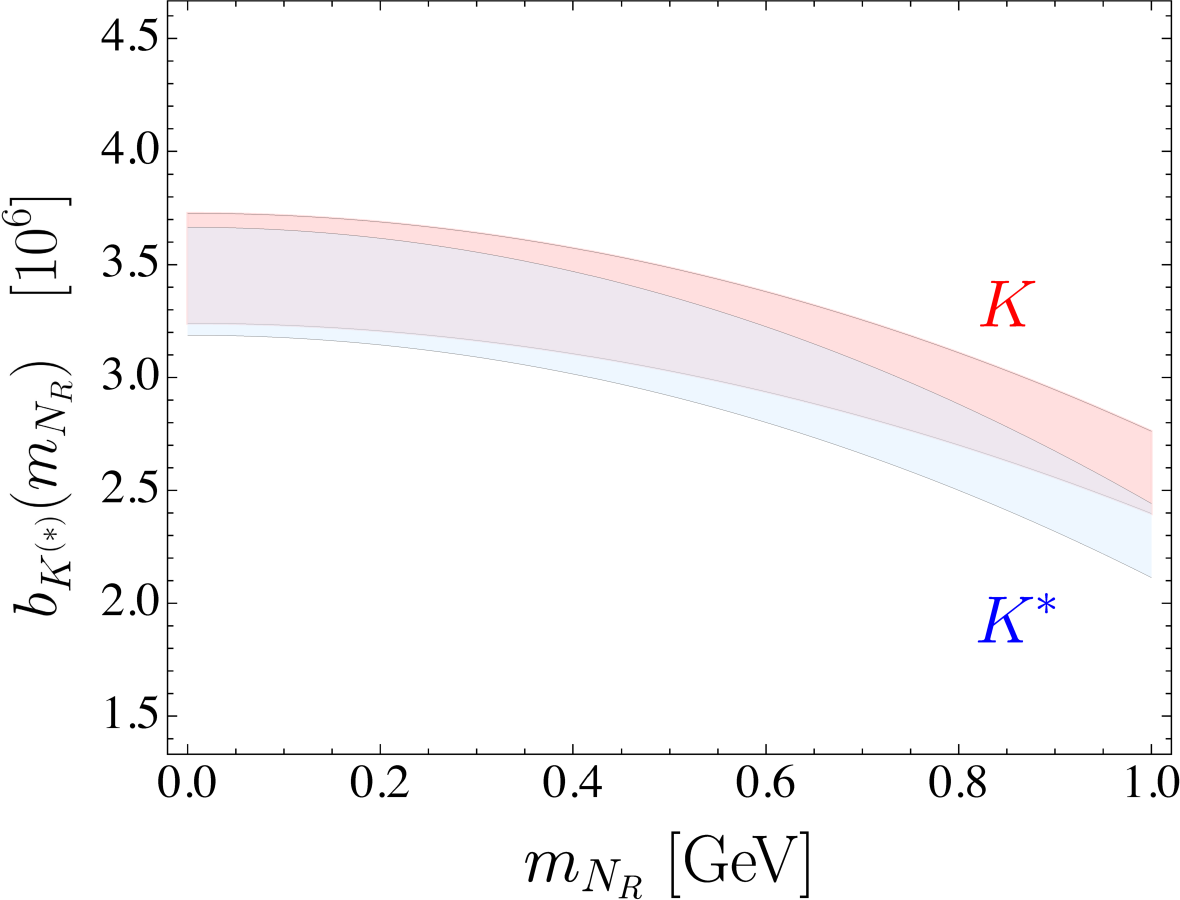}
\caption{\small \sl
Upper and lower plot correspond to the functions $a_{D^{(\ast )}}(m_{N_R}^2)$ and $b_{K^{(\ast )}}(m_{N_R}^2)$ that appear in Eq.~\eqref{eq:rR}, to be multiplied by $10^3$ and $10^6$, as indicated in each plot. 
They are obtained by computing the ratios of branching fractions $\mathcal{B}(B \to D^{(\ast )} \tau N_R)/\mathcal{B}(B \to D^{(\ast )} \tau\bar\nu)^\mathrm{SM}$ and $\mathcal{B}(B \to K^{(\ast )} N_RN_R)/\mathcal{B}(B \to K^{(\ast )} \nu\bar\nu)^\mathrm{SM}$, up to the coupling to the right-right operators $ |C_{RR}|^2$ and $|\widetilde C_{RR}|^2$, respectively.}
\label{fig:mNR-functions} 
\end{figure}

Clearly, when computing the decay rates involving the above operators, there is no interference with the SM contributions, and we have:  
\begin{align}
\mathcal{B}(B\to D^{(\ast )}\tau \text{`inv'}) &= \mathcal{B}(B\to D^{(\ast )} \tau\nu)^\mathrm{SM}\cr
                      &+ \mathcal{B}(B\to D^{(\ast )} \tau N_R)\,,\cr 
\mathcal{B}(B\to K^{(\ast )} \text{`inv'})  &=  \mathcal{B}(B \to K^{(\ast )} \nu\bar\nu)^\mathrm{SM}\cr
& +\mathcal{B}(B \to K^{(\ast )} N_R N_R)\,,
\end{align}
where the missing energy, corresponding to {\em invisible} particles (`inv'), is attributed to neutrinos in the SM and to the RH neutrino ($N_R$) in the scenario considered here. Note that we consider one species of $N_R$, but in principle, we can have $3$ of them without running into any problem with phenomenology. 
We computed the decay rates using \eqref{eq:RR}, see also Refs.~\cite{Felkl:2023ayn,Datta:2022czw}, and the numerical results can be summarized in:
\begin{align}
\label{eq:rR}
r_{D^{(\ast )}}  &=1+ \left|C_{RR}\right|^2\, a_{D^{(\ast )}}(m_{N_R}^2),\nonumber\\
R_{K^{(\ast )}}^{\text{`inv'}}  &=1+ \left|\widetilde C_{RR}\right|^2\, b_{K^{(\ast )}}(m_{N_R}^2) ,
\end{align}
where the functions $a_{D^{(\ast )}}(m_{N_R}^2)$ and $b_{K^{(\ast )}}(m_{N_R}^2)$  
and are shown in Fig.~\ref{fig:mNR-functions}. In this work we focus on the range $0\leq m_{N_R}\leq 1$~GeV, common to all the channels, namely, 
$m_{N_R} \leq m_B-m_{D^{(\ast )}}-m_\tau$ and $m_{N_R} \leq (m_B-m_{K^{(\ast )}})/2$.
The experimental results \eqref{RDexp} and \eqref{RKexp} can now be converted to $|C_{RR}|$ and $|\widetilde C_{RR}|$ for each $m_{N_R}$.

\paragraph*{\underline{$b\to c\tau  \text{`inv'}$}}\label{par:bctaunu}: With $|C_{RR}|$ extracted in a way described above, one can test our proposal by computing the angular observables of $B\to D^{(\ast )}\tau   \text{`inv'}$, see e.g.~\cite{Murgui:2019czp,Bernlochner:2024xiz}, and then compare them to their SM estimates. We choose to consider the forward-backward asymmetry ($A_\mathrm{fb}^{D^{(\ast )}}$), the $\tau$-polarization asymmetry ($P_\tau^{D^{(\ast )}}$) and the fraction of the longitudinally polarized $D^\ast$ emerging from  $B\to D^\ast \tau \text{`inv'}$ ($F_L^{D^\ast}$). In Tab.~\ref{tab:1}, we provide our results for these quantities in the SM and in our scenario described by Eq.~\eqref{eq:RR}. For illustrative purposes, we choose three different values of $m_{N_R}$, and besides $0$ and $1$~GeV, we use $m_{N_R}=0.6$~GeV, which appears to be the optimal option for explaining the binned experimental $B\to K\nu\bar\nu$ data~\cite{Bolton:2024egx,Fridell:2023ssf}. 
\begin{table}[!t]
\renewcommand{\arraystretch}{1.8}
\centering
\begin{tabular}{|c|c|ccc|}
\hline 
Quantity                &  SM &     Case 1.  & Case 2. & Case 3. \\ \hline 
$ | C_{RR}| \times 10^2$  &    $ -$   &   $1.6(2)$         &    $2.0(2)$           &    $3.1(4)$     \\  \hline 
$A_\mathrm{fb}^{D}$      &  $0.360(0)$       &   $0.360(0)$    & $0.341(4)$        &    $0.329(4)$     \\ 
$A_\mathrm{fb}^{D^{\ast }}$&    $-0.06(1)$    &   $-0.06(1)$     &   $-0.06(1)$         &   $-0.06(1)$    \\ 
$P_\tau^{D}$         &   $0.325(3)$     &   $0.25(2)$         &    $0.26(2)$           &    $0.28(1)$     \\ 
$P_\tau^{D^{\ast }}$ &   $-0.51(2)$     &   $-0.39(4)$         &    $-0.41(3)$           &    $-0.43(3)$    \\ 
$F_L^{D^{\ast }}$&  $0.46(1)$      &   $0.46(1)$         &    $0.46(1)$           &    $0.45(1)$     \\ 
$R_{B_c}$&     $1$   &   $1.17(10)$         &    $1.29(13)$           &    $1.63(31)$     \\ 
$R_{J/\psi}$&     $0.258(4)$   &   $0.296(10)$         &    $0.292(10)$           &    $0.277(7)$     \\ 
\hline
\end{tabular}
\caption{ \sl \small $| C_{RR}|$ values are extracted from $R_{D^{(\ast )}}^\mathrm{exp}$ in three different cases corresponding to three different values of the RH neutrino mass, in increasing order, $m_{N_R} \in \{0,\, 0.6,\, 1\}$~GeV. Those results are then used to compute the observables discussed in the text. The SM results are also shown for comparison. Of all the quantities listed above, only a few have been studied experimentally so far: $P_\tau^{D^{\ast }}=-0.37(54)$~\cite{Belle:2017ilt}, $F_L^{D^\ast} = 0.60(9)$~\cite{Belle:2019ewo}, $0.43(7)$~\cite{LHCb:2023ssl}, $R_{J/\psi}=0.71(25)$~\cite{LHCb:2017vlu}. Expressions for the observables in the table can be found in Appendix \ref{sec:b2c}}
\label{tab:1} 
\end{table}
As it can be seen from Tab.~\ref{tab:1}, the only appreciable difference with respect to the SM is found for the $\tau$ polarization asymmetries ($P_\tau^{D^{(\ast )}}$), the absolute values of which are reduced by about $20\%$. Furthermore, $\mathcal{B}(B_c\to \tau  \text{`inv'})= \mathcal{B}(B_c\to \tau \nu)^\mathrm{SM}+ \mathcal{B}(B_c\to \tau N_R)$, becomes moderately larger than in the SM, which in Tab.~\ref{tab:1} is presented in terms of $R_{B_c}=\mathcal{B}(B_c\to \tau  \text{`inv'})/\mathcal{B}(B_c\to \tau  \nu)^\mathrm{SM}$. In other words, the experimental test of the scenario proposed here can only go through high precision, both in hadronic quantities and in experimental studies. To close this discussion we should also mention that $R_{J/\psi}=  \mathcal{B}(B_c\to J/\psi \tau \nu)/\mathcal{B}(B_c\to J/\psi \mu \nu)$, which is similar to $R_{D^{(\ast})}$, can be predicted thanks to the accurate results for the form factors computed using lattice QCD~\cite{Harrison:2020gvo}, cf. Tab.~\ref{tab:1}. One could also calculate different baryonic quantities, most notably, the ratio $R_{\Lambda_c}=\mathcal{B}(\Lambda_b \rightarrow \Lambda_c \tau \nu)/\mathcal{B}(\Lambda_b \rightarrow \Lambda_c \ell \nu)$. The formulas for semileptonic baryon decays in the case of massive right-handed neutrino are not yet available in the literature and would deserve a separate study.

\paragraph*{\underline{$b\to s  \text{`inv'}$}}: In this case, there are fewer observables that would allow us to check on the viability of this model. 
We extract $|\widetilde C_{RR}(m_{N_R})|$ from the comparison between \eqref{eq:rR} and $R_K^{ \text{`inv'}\ \mathrm{(exp)}}$ given in \eqref{RKexp}. 
We then use $|\widetilde C_{RR}|$ to check on the fraction of the longitudinally polarized $K^\ast$ ($F_L^{K^\ast}$), and predict the $\mathcal{B}(B_s\to  \text{`inv'})$ which is completely negligible in the SM, while here, due to the mass of $N_R$ it can be as large as $\mathcal{O}(10^{-6})$. 
Finally, as one could expect, the enhancement of $R_{K}^{ \text{`inv'}\ \mathrm{(exp)}}$ accommodated by this model implies a similar enhancement of $R_{K^\ast}^{ \text{`inv'} }$ and of  $R_{D_s}^\text{`inv'}=\mathcal{B}(B_c\to D_s \text{`inv'})/\mathcal{B}(B_c\to D_s \nu\bar\nu)^\mathrm{SM}$. For the latter case, there are no experimental attempts yet, but the lattice QCD results of the corresponding form factors already exist~\cite{Cooper:2021bkt}. 
Regarding $R_{K^\ast}^{ \text{`inv'}}$, we obtain the increase similar to $R_{K}^{ \text{`inv'}}$. Notice, however, that the experimental bounds of Ref.~\cite{Belle:2017oht} indicate that $R_{K^\ast}^{ \text{`inv'}} < 2.7$ and $R_{K}^{ \text{`inv'}} < 3.6$. Since the  Belle~II value for $R_{K}^{ \text{`inv'}}$ turned out to be larger than that bound, cf. Eq.~\eqref{RKexp}, we believe it is judicious to wait for Belle~II to assess $R_{K^\ast}^{ \text{`inv'}}$ experimentally.
Our results for the same three masses $m_{N_R}$ are provided in Tab.~\ref{tab:2}.

\begin{table}[!h]
\renewcommand{\arraystretch}{1.8}
\hspace*{-3mm}
\begin{tabular}{|c|c|ccc|}
\hline 
Quantity                                          &  SM & Case 1.  & Case 2. & Case 3. \\ \hline 
$ \left| \widetilde C_{RR}\right| \times 10^3$  &    $ -$   &   $1.1(2)$         &    $1.2(2)$           &    $1.3(2)$     \\  \hline 
$R_{K^\ast}^{ \text{`inv'}}$     &   $1$    &  $5.3\pm 1.5$       &   $5.2\pm 1.4$    & $4.9\pm 1.3$             \\ 
$F_L^{K^\ast}$      &  $0.48(7)$       &   $0.47(7)$    & $0.47(7)$        &    $0.47(7)$     \\ 
$\mathcal{B}(B_s\to  \text{`inv'})$  &    $0$    &   $0$     &   $(9\pm 3) \times 10^{-7}$         &   $(3\pm 1)\times  10^{-6}$   \\ 
$R_{D_s}^\text{`inv'}$  &    $1$    &   $5.3\pm 1.4$     &   $5.4\pm 1.4$         &   $5.5 \pm 1.4$   \\ 
\hline
\end{tabular}
\caption{ \sl \small $ \left| \widetilde C_{RR}\right|$ is extracted from $R_K^{ \text{`inv'}\ \mathrm{(exp)}}$ in three cases, i.e. three different values of the RH neutrino mass (cf. caption of Tab.~\ref{tab:1}), and then used to evaluate the observables for which there is no experimental information yet. Expressions for the observables in the table can be found in Appendix \ref{sec:b2s}}
\label{tab:2} 
\end{table}

\section{Concrete Model}
\begin{figure}[t!]
\centering
\includegraphics[width=.85\linewidth]{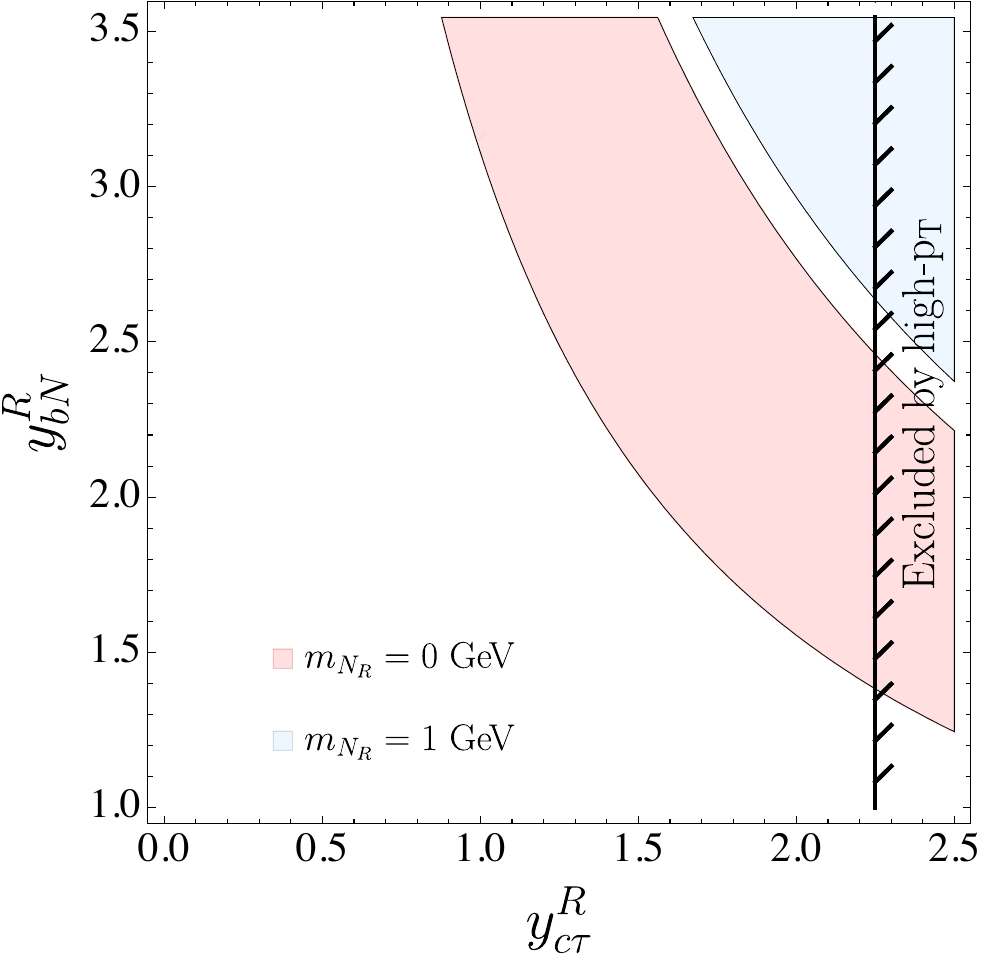}\\
\includegraphics[width=0.85\linewidth]{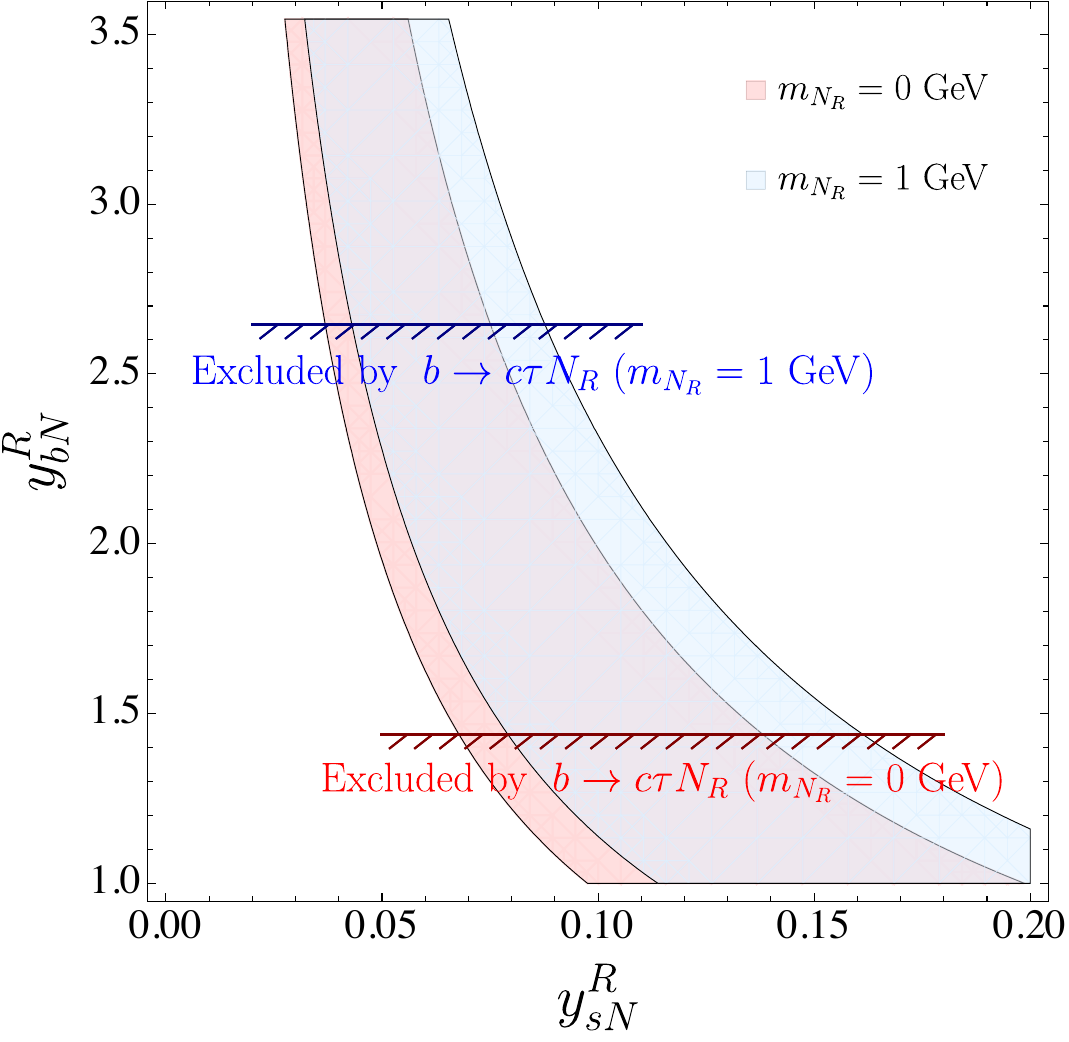}
\caption{\small \sl
In the upper plot we show $y^R_{b N}$ Vs $y^R_{c\tau}$ as obtained from the requirement of $2\sigma$ consistency with $R_{D^{(\ast )}}^\mathrm{exp}$, for two values of $m_{N_R}$. Similarly, in the lower one we show $y^R_{b N}$ Vs $y^R_{s N}$ obtained from $\widetilde C_{RR}(m_{N_R})$ deduced from the consistency with $R_K^{\nu\nu\ \mathrm{(exp)}}$. The lower limit on $y^R_{b N}$ from the upper plot is indicated in the lower one. }
\label{fig:Yukawas} 
\end{figure}

We now illustrate the scenario discussed above with a concrete model. For that purpose, for the needed heavy mediator, we will assume that that there is only one  $\mathcal{O}(1\,\mathrm{TeV})$ scalar leptoquark. 
Since we need it to have RH couplings to both quarks and leptons, we can opt for either a weak singlet or triplet. After inspection of the complete list of Lagrangians~\cite{Dorsner:2016wpm}, we see that the scalar leptoquark $S_1$ indeed couples to RH quarks and leptons, including the RH neutrino~\cite{Azatov:2018kzb}. Note that in terms of the SM quantum numbers, color triplet, weak singlet, and the hypercharge $-1/3$,  $S_1$ is denoted by $(\bar 3, 1, -1/3)$, and its Lagrangian contains the following RH interactions:
\begin{equation}
\label{eq:model}
\mathcal{L} \supset y^R_{c \tau}\,\overline{c^c} P_R \tau\, S_1 + y^R_{sN}\,\overline{s^c} P_R   N_R \, S_1  +  y^R_{b N}\, \overline{b^c} P_R   N_R \, S_1 + \mathrm{h.c.}\,,
\end{equation}
in an obvious notation, where $y^R_{c \tau}$, $y^R_{b N}$ and $y^R_{s N}$ are the three non-zero Yukawa couplings, which together with $m_{S_1}$ amount to four model parameters. We know that the current experimental bound on $m_{S_1}\gtrsim 1.5$~TeV~\cite{Angelescu:2021lln}, which we will set to $m_{S_1}=2$~TeV in the following. At the low energy scales, after integrating out $S_1$, one indeed retrieves the Lagrangian~\eqref{eq:RR}, with the Wilson coefficients expressed in terms of the $S_1$-model parameters as:
\begin{equation}
\label{eq:Wils2}
C_{RR}  = - \frac {v^2}{4 m_{S_1}^2 } y^{R*}_{c \tau} y^R_{b N}\,, \quad
\widetilde C_{RR}   = - \frac {v^2}{2 m_{S_1}^2 } y^{R*}_{s N} y^R_{b N}\,,
\end{equation}
where $v^2=(\sqrt{2} G_F)^{-1}$. To make the plots less crowded, we now restrict our attention to $m_{N_R}= 0$ and $m_{N_R}= 1$~GeV. We interpret the values of $C_{RR}(m_{N_R})$ and $\widetilde C_{RR}(m_{N_R})$, extracted from the requirement of consistency with experimental data in the previous Section, cf. Tabs.~\ref{tab:1} and~\ref{tab:2}, in terms of the mentioned Yukawa couplings. The corresponding results are shown in Fig.~\ref{fig:Yukawas}.  
While $y^R_{sN}$ always remains small, the coupling $y^R_{bN}$ can be large and covers the range of values from about $1.3$ to the perturbativity limit $\sqrt{4\pi}$. The coupling $y^R_{c\tau}$, instead, cannot take large values as it is bounded from above utilizing the constraints derived from experimental studies of the high-$p_T$ tails of differential $p p \to \tau\tau$(+\text{\emph{soft jets}}) cross section~\cite{Allwicher:2022gkm}. Similar analysis of the $p p \to \tau\nu$ channel, that does not interfere with the SM process, results in a very weak bound on $y_{bN}^R$ that does not impact the phenomenological analysis. In High-$p_T$ package RH neutrinos are not implemented and in order to extract the bounds we have exploited the similarity between $S_1$ and $S_3$ leptoquarks.
The signature of a $S_1$-mediated process, e.g. $\overline{c}_R b_R \to \overline{\tau}_R N_R$, is similar to the $S_3^{1/3}$-mediated one $\overline{c}_L b_L \to \overline{\tau}_L \nu_L$, and it suffices to analyse $S_3$ cross section and remove the interference term of $S_3$ amplitude with the SM to obtain the bound on $y_{c\tau}^R y_{bN}^R$.

All of the phenomenology discussed in the previous section obviously applies here also. The only difference is that for $m_{N_R} \leq 0.19$~GeV, this model can contribute to the leptonic decay of $D_s$ as  $\mathcal{B}(D_s \to \tau \text{`inv'}) = \mathcal{B}(D_s \to \tau \nu)^\mathrm{SM}+\mathcal{B}(D_s \to \tau N_R)$. We find that such a contribution is completely negligible (owing to the fact that  $y^R_{c\tau}y^R_{sN}$ is small). We also checked that $y_{c\tau}^R$ and $y_{bN}^R$ give tiny contributions to $\mathcal{B}(Z\rightarrow \tau \bar\tau)$ and $\mathcal{B}(Z\rightarrow  \text{`inv'})$, respectively. Furthermore, and prompted by Ref.~\cite{Aloni:2017eny}, we checked that the contributions from this model to the decays of quarkonia, such as $\mathcal{B}(\psi (nS) \to \tau\bar\tau)$, $\mathcal{B}(\Upsilon (nS) \to N_R N_R)$ or $\mathcal{B}(\phi \to N_R N_R)$, are completely negligible too.

 Finally, one needs to verify the consistency of our proposal with the $B_s-\overline B_s$ mixing. Since the relevant hadronic matrix element is the same as in the SM, namely, 
$\langle \overline B_s \vert (\bar s \gamma_\mu P_R b)(\bar s \gamma_\mu P_R b)\vert B_s\rangle =\langle \overline B_s \vert (\bar s \gamma_\mu P_L b)(\bar s \gamma_\mu P_L b)\vert B_s\rangle$,
one can simply rescale $\Delta m_{B_s}  =(1 + C_{S_1}/C_{SM}) \Delta m_{B_s}^\mathrm{SM}$, after identifying 
\begin{equation}
C_{S_1} = \frac{   \left( y_{sN}^{R \,\ast}\, y_{bN}^R\right)^2}{256 \pi^2 \lambda_t^2  }  \frac{ v^2 }{m_{S_1}^2} = \frac{  \widetilde C_{RR}^2 }{ 64 \pi^2 \lambda_t^2} \frac{m_{S1}^2}{v^2}. 
\end{equation}
Unfortunately, the value of $\Delta m_{B_s}^\mathrm{SM}$ depends on two input quantities which are not well established: (i) $\lambda_t$, that is related to $|V_{cb}|$ via the CKM unitarity inheriting the well-known disagreement between $|V_{cb}|$ extracted from the inclusive and from the exclusive semileptonic decays (see  \cite{Becirevic:2023aov}), (ii) FLAG result for the lattice QCD value of the above hadronic matrix element~\cite{FlavourLatticeAveragingGroupFLAG:2021npn} is about $2\sigma$ larger than the one obtained in Ref.~\cite{Cooper:2020wnj} in which $N_f=2+1+1$ quark flavors have been included in simulations with the staggered quark action. If the inclusive value of $|V_{cb}|$ is confirmed through exclusive decays, and if the result of Ref.~\cite{Cooper:2020wnj} is confirmed in simulations with $N_f=2+1+1$ sea quark flavors (but by using a different lattice QCD action), then the very accurate $\Delta m_{B_s}^\mathrm{exp}=17.765(6)\,\mathrm{ps}^{-1}$~\cite{ParticleDataGroup:2024cfk} could be used to compromise the validity of this model. However, given the current ambiguity of $V_{cb}$ and the mixing matrix element, we cannot draw any firm conclusions.

\section{Summary}\label{sec:summary}
Finding a BSM scenario that could simultaneously accommodate two recently observed deviations from the SM expectations, namely,  $r_{D^{(\ast )}} = R_{D^{(\ast )}}^\mathrm{exp}/R_{D^{(\ast )}}^\mathrm{SM}> 1$, $R_{K}^{ \text{`inv'}\,\mathrm{(exp)}}> 1$, is quite complicated, if one stick to the effective theory with a minimal number of parameters, that can be scrutinized experimentally to either validate or refute the model. In this paper, we find that one can build such a scenario if we allow only RH couplings of the heavy mediator to both quarks and leptons. When computing the contribution of such a model to various physical quantities, there is no interference with the SM. As for the particle content at low energy, we assume the existence of the neutral RH lepton $N_R$ (RH neutrino) of $m_{N_R} \lesssim 1$~GeV, without trying to explain its origin. Such a BSM model can encapsulate physics up to $\mathcal{O}(1~\mathrm{TeV})$ scales and describe the flavor physics discrepancies mentioned above. We discussed the phenomenology of such a scenario and listed several observables that can be used to test it, but we found that this can be possible only through $R_{K^\ast}^{ \text{`inv'}}$ and high precision both in experiment and in theory (mainly by taming the hadronic uncertainties). Such is the case with oscillation frequency of the $B_s-\overline B_s$ system ($\Delta m_{B_s}$), which is very precisely established experimentally but for which a similar theoretical uncertainty is still missing.  
We also illustrate the scenario we propose on an explicit model with a minimal set of parameters and with a singlet scalar leptoquark ($S_1$) as a heavy mediator. 

\section*{Acknowledgments}
 S.F., N.K. and L.P. acknowledge financial support from the Slovenian Research Agency (research core funding No. P1-0035 and N1-0321).
This project has received support from the European Union’s Horizon 2020 research and innovation program under the Marie Sklodowska-Curie grant agreement No 860881-HIDDeN.

\appendix

\section{Expressions relevant to $b\to s N_RN_R$ }\label{sec:b2s}

The expressions for $\mathcal{B}(B\to K^{(\ast )}\nu\nu)^\mathrm{SM}$ can be found in Refs.~\cite{Buras:2014fpa,Becirevic:2023aov}. 
Here we provide those relevant to our scenario in which the $b \to s \text{`inv'}$ processes are described by the following effective Lagrangian:
\begin{align}
	\label{eq:RR}
	\mathcal{L}^{b\to s N_R N_R} =& - \sqrt 2 G_F \widetilde C_{RR} ( \bar s \gamma_\mu P_R b ) (\bar N_R \gamma^\mu P_R N_R) +\mathrm{h.c.}\nonumber\,.
\end{align}
We use the standard decomposition of the hadronic matrix elements:
\begin{equation}
	\label{eq:ffKst}
	\begin{split}
		\langle {K}(k) | \bar{s}\gamma^\mu  b | \bar{B}(p) \rangle &= \Big{[}(p+k)^\mu- \dfrac{m_B^2-m_K^2}{q^2}q^\mu\Big{]} f_{+}(q^2)  \\ 
		&\quad+ \dfrac{m_B^2-m_K^2}{q^2} q^\mu f_0(q^2)\,,
	\end{split}
\end{equation}
\begin{align}
	\label{eq:BKst-ff}
	\langle {K}^\ast(k) | \bar{s}\gamma_\mu (1+&\gamma_5) b | \bar{B}(p) \rangle = \varepsilon_{\mu\nu\rho\sigma} \varepsilon^{\ast\nu}p^\rho k^\sigma \dfrac{2 V(q^2)}{m_B+m_{K^\ast}}\nonumber\\[0.4em]
	&+i\varepsilon_\mu^\ast(m_B+m_{K^\ast})A_1(q^2) \nonumber\\
	&- i(p+k)_\mu (\varepsilon^\ast\cdot q)\dfrac{A_2(q^2)}{m_B+m_{K^\ast}}\\
	&-i q_\mu (\varepsilon^\ast\cdot q) \dfrac{2 m_{K^\ast}}{q^2} \left[A_3(q^2)-A_0(q^2)\right]\,, \nonumber
\end{align}
where $\varepsilon_\mu$ is the polarization vector of $K^\ast$, and  $f_{+}(q^2)$, $f_{0}(q^2)$, $V(q^2)$ and $A_{0,1,2,3}(q^2)$ are the form factors. Besides the form factor $A_3(q^2)$, already present in the above definition, we also introduce  $A_{12}(q^2)$ for convenience. They are related to $A_1(q^2)$ and $A_2(q^2)$ as: 
\begin{align}
	A_3(q^2) &= \dfrac{m_B+m_{K^\ast}}{2 m_{K^\ast}}A_1(q^2)
	- \dfrac{m_B-m_{K^\ast}}{2 m_{K^\ast}}A_2(q^2), \nonumber \\[0.4em]
	A_{12}(q^2) &= \dfrac{(m_B+m_{K^\ast}) (m_B^2 -m_{K^\ast}^2 -q^2)}{16 m_B m_{K^\ast}^2} A_1(q^2) \nonumber\\
	&- \dfrac{\lambda_{BK^\ast}}{16 m_B m_{K^\ast}^2 (m_B+m_{K^\ast}) } A_2(q^2)\,.
\end{align}
At $q^2=0$ they satisfy the relations: $A_3(0)=A_0(0)$ and $8 m_B m_{K^\ast} A_{12}(0) = (m_B^2 - m_{K^\ast}^2) A_0(0)$. 
For shortness, we write $ \lambda_{BK^{(\ast )}} \equiv \lambda(\sqrt{q^2},m_B,m_{K^{(\ast )}})$.~\footnote{$\lambda (x,y,z) = \left[ x^2 - (y+z)^2\right]\,  \left[ x^2 - (y-z)^2\right]$, is used throughout.}
We emphasize once again that contributions of our model do not interfere with the SM amplitude, and we have 
\begin{equation}
	\begin{split}
		\mathcal{B}(B\to K^{(\ast )} \text{`inv'})  &=  \mathcal{B}(B \to K^{(\ast )} \nu\bar\nu)^\mathrm{SM}\cr
		&\quad +\mathcal{B}(B \to K^{(\ast )} N_RN_R)\,,
	\end{split}
\end{equation}
where $\text{`inv'}$ stands for the missing energy, attributed to neutrinos in the SM, and to the pair of RH neutrinos ($N_R$) in our model. 
The expression for the latter term reads:
\begin{equation}
	\begin{split}
		&\frac{d\mathcal{B}(B \to K N_RN_R) }{dq^2} =\vert \widetilde C_{RR}\vert^2 \times \cr
		&\qquad\times \biggl[ a_+^K(q^2)  \vert f_+(q^2)\vert^2 + a_0^K(q^2)  \vert f_0(q^2)\vert^2\biggr],
	\end{split}
\end{equation}
with
\begin{equation}
	\begin{split}
		a_+^K(q^2)  &= b_R^K(q^2)\, \lambda_{BK}  \left(1-\frac{m_{N_R}^2}{q^2}\right),\cr
		a_0^K(q^2)  &= b_R^K(q^2)\, (m_B^2-m_K^2)^2\,\frac{ 3 m_{N_R}^2}{q^2} ,
	\end{split}
\end{equation}
and
\begin{align}\label{eq:bRK}
	b_R^K(q^2) &= \frac{\tau_B G_F^2}{192 \pi^3 m_B^3} \sqrt{\lambda_{BK}} \, \sqrt{1 - \frac{4 m_{N_R}^2}{q^2}} \,.
\end{align}
Clearly the term proportional to $|f_0(q^2)|^2$ is absent in the massless neutrino limit.

In the case of the vector meson in the final state we have:
\begin{align}\label{eq:BKst}
	&\frac{d\mathcal{B}(B \to K^\ast N_RN_R) }{dq^2} =\vert \widetilde C_{RR}\vert^2 \times \biggl[  a_V^{K^\ast}(q^2)  \vert V(q^2)\vert^2 + \biggr. \cr
	&\biggl.  a_1^{K^\ast}(q^2)  \vert A_1(q^2)\vert^2 + a_{12}^{K^\ast}(q^2)\vert A_{12}(q^2)\vert^2 + a_0^{K^\ast}(q^2)  \vert A_0(q^2)\vert^2 \biggr],
\end{align}
where
\begin{align}
	a_V^{K^\ast}(q^2)  &= b_R^{K^\ast}(q^2)\,\frac{2 q^2 \lambda_{B K^\ast } }{ (m_B+m_{K^\ast})^2} \left(1-\frac{m_{N_R}^2}{q^2}\right),\cr
	a_1^{K^\ast}(q^2)  &= b_R^{K^\ast}(q^2)\, 2 q^2  (m_B+m_{K^\ast})^2  \left(1-\frac{m_{N_R}^2}{q^2}\right),\cr
	a_{12}^{K^\ast}(q^2)  &= b_R^{K^\ast}(q^2)\, 64\, m_B^2 m_{K^\ast}^2  \left(1-\frac{m_{N_R}^2}{q^2}\right),\cr
	a_{0}^{K^\ast}(q^2)  &= b_R^{K^\ast}(q^2)\, \lambda_{B K^\ast }\,\frac{ 3 \, m_{N_R}^2}{q^2},
\end{align}
and $b_R^{K^\ast}(q^2)$ is obtained from Eq.~\eqref{eq:bRK} after replacing $m_K$ by $m_{K^\ast}$.
The above way to write the decay rate is useful because it separates the longitudinally ($\varpropto |A_0(q^2)|^2$ and $|A_{12}(q^2)|^2$) 
from the transversely polarized parts ($\varpropto |A_1(q^2)|^2$ and $|V(q^2)|^2$) of the decay. In that way 
\begin{align}\label{eq:BKstL}
	&\frac{d\mathcal{B}_L }{dq^2} =\vert \widetilde C_{RR}\vert^2 \, \biggl[  a_{12}^{K^\ast}(q^2)\vert A_{12}(q^2)\vert^2 + a_0^{K^\ast}(q^2)  \vert A_0(q^2)\vert^2 \biggr],
\end{align}
so that
\begin{equation}
	F_L = \frac{ \mathcal{B}_L ( B \to K^\ast \text{`inv'})  }{  \mathcal{B} ( B \to K^\ast \text{`inv'}) }\,,
\end{equation}
where the numerator and denominator are obtained after integrating Eqs.~(\ref{eq:BKst},\ref{eq:BKstL}) between $q^2_\mathrm{min}= 4 m_{N_R}^2$ and $q^2_\mathrm{max}= (m_B -  m_{K^\ast})^2$, and added to the respective SM contributions. 
Again, the term proportional to $|A_0(q^2)|^2$ vanishes in the massless neutrino limit.
Note that the calculation of the $B_c \to D_s N_R N_R$ spectrum precisely follows the above procedure, with appropriate changes in form factors~\cite{Cooper:2021bkt} and numerical inputs.

Finally, one of the channels that is particularly interesting in the case of massive (RH) neutrinos is $B_s\to \text{`inv'}$. Its decay rate in our model reads:
\begin{equation}
	\begin{split}
		\mathcal{B}(B_s \to  N_RN_R)  &= \frac{ \tau_{B_s}G_F^2  }{16 \pi}  \vert \widetilde C_{RR}\vert^2 m_{B_s} f_{B_s}^2 \times \cr
		&\qquad \times m_{N_R}^2 \sqrt{ 1- \frac{4 m_{N_R}^2}{m_{B_s}^2}}\, , 
	\end{split}
\end{equation}
where $f_{B_s}$ is the decay constant that parametrizes the relevant hadronic matrix element:
\begin{equation}\label{eq:fBs}
	\langle 0\vert \bar s  \gamma_\mu \gamma_5 b  \vert B_s(p)\rangle = i f_{B_s} p_\mu\,.
\end{equation}

\section{Expressions relevant to $b\to c\tau N_R$ }\label{sec:b2c}

We now turn to the other part of our proposal, i.e. the processes related to $b\to c \tau N_R$, described by 
\begin{align}
	\label{eq:RR2}
	\mathcal{L}^{b\to c \tau N_R} 
	=& - \sqrt 2 G_F \,C_{RR} ( \bar c \gamma_\mu P_R b ) (\bar \tau  \gamma^\mu P_R N_R)+\mathrm{h.c.}\,.\nonumber
\end{align}

\subsection{$\mathcal{B}(B_c \to  \tau N_R)$}

The simplest to consider is the leptonic decay $B_c\to \tau\text{`inv'}$. Besides the SM contribution to this mode, $\mathcal{B}(B_c\to \tau \nu)^\mathrm{SM}$, by using the above Lagrangian, we obtain
\begin{equation}
	\begin{split}
		\mathcal{B}(B_c \to  \tau N_R)  
		& =  \frac{ \tau_{B_c}G_F^2  }{8 \pi m_{B_c}^2 }  \vert  C_{RR}\vert^2 f_{B_c}^2 \lambda^{1/2}(m_{B_c},m_\tau,m_{N_R}) \cr
		& \times \biggl[  m_{B_c}^2 (m_\tau^2 +  m_{N_R}^2) - (m_\tau^2 -  m_{N_R}^2)^2\biggr] \,,
	\end{split}
\end{equation}
where $f_{B_c}$ is the decay constant defined in a way analogous to Eq.~\eqref{eq:fBs}.

\subsection{$B\to D^{(*)} \tau N_R$}

The hadronic matrix elements $\langle {D}(k) | \bar{s}\gamma^\mu  b | \bar{B}(p) \rangle$ and  $\langle {D}^\ast(k) | \bar{s}\gamma_\mu (1+\gamma_5) b | \bar{B}(p) \rangle$ are defined in the same way as for the $B\to K^{(*)}$ case in the equation~\eqref{eq:ffKst} with appropriate mass and form factor replacements (the discussion on form factors in $B\to D^{(*)}$ can be found in~\cite{Becirevic:2024pni}). 


Once again, one have to keep in mind the relation
\begin{equation}
	\begin{aligned}
		\mathcal{B}(B\to D^{(\ast )}\tau \text{`inv'})  &=  \mathcal{B}(B \to D^{(\ast )} \tau \bar\nu)^\mathrm{SM}\cr
		&\quad +\mathcal{B}(B \to D^{(\ast )} \tau N_R)\,.
	\end{aligned}
\end{equation}
While the Standard Model contribution is known~\cite{FlavourLatticeAveragingGroupFLAG:2021npn}, for the new physics contribution we use leptonic matrix elements for the massive right-handed neutrino defined as in~\cite{Datta:2022czw} to obtain the expression for the angular differential distribution for $B\to D \tau \nu$:
\begin{equation}
	\label{eq:diff_ang_D}
	\begin{split}
		\frac{d^2 \mathcal{B}(B\to D \tau N_R)}{dq^2 \,d \cos{\theta_{\tau}}}&=a(q^2,m_{N_R})+b(q^2,m_{N_R})\cos{\theta_{\tau}}\\
		&\quad+c(q^2,m_{N_R})\cos^2{\theta_{\tau}}\,.
	\end{split}
\end{equation}
Here the coefficients $a,b,c$ have the following values:
\begin{equation}
	\label{eq:ang_coeff_D}
	\begin{aligned}
		a /\tilde{N}&= 4  \left[ \left| f_+ (q^2) \right| ^2\frac{\lambda_{BD}}{q^2} (q^2  - m_{\tau}^2 - m_{N_R}^2) +\right.  \\ 
		&\hspace{2.5em} \left| f_0 (q^2) \right| ^2\frac{(m_B^2-m_D^2)^2}{q^2}\times \\
		&\hspace{4em}\left. \left((m_{\tau}^2+m_{N_R}^2)-\frac{(m_{\tau}^2-m_{N_R}^2)^2}{q^2}\right)\right]\,,\\
		b /\tilde{N}&= 8\,  \re{\left( f_+(q^2) f_0^*(q^2)\right)}\times\\
		&\hspace{4em}\frac{\sqrt{\lambda_{BD}\lambda_{\tau N}}(m_B^2-m_D^2)(m_{\tau}^2-m_{N_R}^2)}{q^4}\,,\\
		c /\tilde{N}&= -4    \left| f_+ (q^2) \right| ^2\frac{\lambda_{BD} \lambda_{\tau N}}{q^4}\, ,
	\end{aligned}
\end{equation}
where $\lambda_{\tau N}$ is abbreviation for $\lambda (\sqrt{q^2},m_{N_R} , m_{\tau})$, and $\lambda_{BD}$ for  $\lambda (\sqrt{q^2},m_D , m_B)$. The common prefactor to all coefficients $\tilde{N}$ is
\begin{equation}
	\tilde{N}(q^2,m_{N_R})= \frac{\tau_B G_F^2 |V_{cb}|^2}{1024 \pi^3 m_B^3 q^2}\sqrt{\lambda_{BD}}\sqrt{\lambda_{\tau N}} |C_{RR}|^2 \,.
\end{equation}
From this equation one can calculate the total branching fraction:

\begin{align}
	\label{eq:BRtot_D}
	&\frac{d\mathcal{B}(B\to D \tau N_R)}{dq^2}= \frac{8\tilde{N}}{3 q^4}\times\\
	\nonumber
	&\left[ \left|f_+(q^2)\right|^2 
	\lambda_{BD} \left(3 q^4- 3 m_{\tau}^2 q^2 - 3 m_{N_R}^2 q^2 -   \lambda_{\tau N} \right) +\right.\\
	&\left. 3 \left| f_0(q^2) \right|^2 \nonumber
	(m_B^2-m_D^2)^2 \left(q^4- m_{\tau}^2 q^2 - m_{N_R}^2 q^2 -   \lambda_{\tau N} \right) \right],
\end{align}
where we add the branching fractions and not the amplitudes since the polarizations of charged leptons are possible to distinguish in the experiment. The total branching fraction is obtained after integrating over $q^2$ from $(m_{\tau}+m_{N_R})^2$ up to $(m_{B}-m_{D})^2$.

Also, the two angular observables were calculated, first of which is forward-backward asymmetry:
\begin{equation}
	\label{eq:AFBD}
	\begin{aligned}
		A_\mathrm{fb}^{D}=&\frac{1}{\mathcal{B}(B\to D\tau \text{`inv'})}\times \\
		&\left(\int dq^2 \left(\int_{0}^{1}-\int_{-1}^{0}\right)d\cos{\theta_{\tau}}\frac{d^2 \mathcal{B}}{dq^2 d\, \cos{\theta_{\tau}}}\right)\\
		=&\frac{1}{\mathcal{B}(B\to D\tau \text{`inv'})}\left(\int d q^2 \,b(q^2,m_{N_R}^2) \right)\,,
	\end{aligned}
\end{equation}
where the coefficient $b$ is given in eq.~(\ref{eq:ang_coeff_D}). The second observable is $\tau$ polarization asymmetry:

\begin{equation}
	\label{eq:Ptau}
	\begin{aligned}
		P_{\tau}^D&= \frac{1}{\mathcal{B}(B\to D\tau \text{`inv'})}\: \int dq^2 \,\left(\frac{d(\mathcal{B}^+ - \mathcal{B}^-)}{dq^2} \right)\\
		&= \frac{1}{\mathcal{B}(B\to D\tau \text{`inv'})}\: \int dq^2\, \frac{8\tilde{N}}{3 q^4}\sqrt{\lambda_{\tau N}}\times\\
		&\quad\left[ 3 \left| f_0(q^2) \right|^2(m_B^2-m_D^2)^2 (m_{N_R}^2-m_{\tau}^2) +\right. \\
		&\qquad\left. \left|f_+(q^2)\right|^2 \lambda_{BD} (2 q^2 + m_{N_R}^2 -m_{\tau}^2) \right]\,,
	\end{aligned}
\end{equation}
where $\mathcal{B}^{\pm}$ denotes branching fractions in which $\tau$ has a specified polarization (see also~\cite{Datta:2022czw}). 
It is important to stress that in the above expressions~\eqref{eq:AFBD}, \eqref{eq:Ptau}, we have omitted the SM contribution in the numerator and kept only the RHN contributions.

The $B\to D^* \tau N_R$ decay, on the other hand, has a richer angular structure:
\begin{equation}
	\begin{aligned}
		\frac{8 \pi}{3}&\frac{d^4 \mathcal{B}(B\to D \tau N_R)}{dq^2d\cos{\theta_{\tau}}d\cos{\theta_D}d\phi}=\\
		&\;\quad\quad(\mathcal{I}_{1s}+\mathcal{I}_{2s}\cos{2\theta_{\tau}}+\mathcal{I}_{6s}\cos{\theta_{\tau}})\sin^2{\theta_D}\\
		&\quad+(\mathcal{I}_{1c}+\mathcal{I}_{2c}\cos{2\theta_{\tau}}+\mathcal{I}_{6c}\cos{\theta_{\tau}})\cos^2{\theta_D}\\
		&\quad+(\mathcal{I}_3 \cos{2\phi}+\mathcal{I}_9 \sin{2\phi}) \sin^2{\theta_D}\sin^2{\theta_{\tau}}\\
		&\quad+(\mathcal{I}_4 \cos{\phi}+\mathcal{I}_8\sin{\phi})\sin{2\theta_D}\sin{2\theta_{\tau}}\\
		&\quad+(\mathcal{I}_5 \cos{\phi}+\mathcal{I}_7\sin{\phi})\sin{2\theta_D}\sin{\theta_{\tau}}\,.
	\end{aligned}
\end{equation}
The angular coefficients $\mathcal{I}_i$ have the following values:
\begin{widetext}
	\begin{equation}
		\begin{aligned}
			\mathcal{I}_{1s}/N&= \left[(m_B+m_{D^*})^2 \left| A_1(q^2)\right|^2+\frac{\lambda_{BD^*}}{(m_B+m_{D^*})^2} \left| V(q^2)\right|^2\right]\frac{1}{q^2}\left(4q^4- 4m_{\tau}^4-4m_{N_R}^4-\lambda_{\tau N} \right) \,,\\
			\mathcal{I}_{2s}/N&= \left[(m_B+m_{D^*})^2 \left| A_1(q^2)\right|^2+\frac{\lambda_{BD^*}}{(m_B+m_{D^*})^2} \left| V(q^2)\right|^2\right]\frac{\lambda_{\tau N}}{q^2}\,,\\
			\mathcal{I}_{1c}/N&= 2 \left[ \frac{64 m_{D^*}^4}{q^2} \left| A_{12}(q^2) \right|^2 \left( 2q^4 - 2 m_{\tau}^2 q^2 -2 m_N^2 q^2 - \lambda_{\tau N}  \right) + \frac{2\lambda_{BD^*}}{q^2} \left| A_0(q^2) \right|^2 \left( q^4 -  m_{\tau}^2 q^2 - m_N^2 q^2 - \lambda_{\tau N} \right) \right] \,,\\ 
			\mathcal{I}_{2c}/N&=-  \frac{64 m_{D^*}^4\lambda_{\tau N}}{q^4} \left| A_{12}(q^2) \right|^2 \,,\\
			\mathcal{I}_{3}/N&=- \left[(m_B+m_{D^*})^2 \left| A_1(q^2)\right|^2-\frac{\lambda_{BD^*}}{(m_B+m_{D^*})^2} \left| V(q^2)\right|^2\right]\frac{ \lambda_{\tau N}}{q^2}\,,\\
			\mathcal{I}_{4}/N&= -\frac{16 m_{D^*}^2 (m_B-m_{D^*})\lambda_{\tau N}}{2q^{3}} \, \re{\left(A_{12}(q^2) A_1^*(q^2)\right)}\,,\\
			\mathcal{I}_{5}/N&= -2 \left[\frac{16 m_{D^*}^2 \sqrt{\lambda_{BD^*}}\sqrt{q^2}}{(m_B-m_{D^*})} \,\re{\left(A_{12}(q^2) V^*(q^2) \right)} + \frac{\sqrt{\lambda_{BD^*}}(m_B+m_{D^*})\left( m_{\tau}^2 - m_N^2  \right)}{\sqrt{q^2}} \,\re{\left(A_0(q^2) A_1^*(q^2)\right)}   \right] \frac{ \sqrt{\lambda_{\tau N}} }{ q^2}\,,\\
			\mathcal{I}_{6s}/N&=   - 8 \sqrt{\lambda_{BD^*}\lambda_{\tau N}} \,\re{\left(A_1(q^2) V^*(q^2)\right)}  \,, \\
			\mathcal{I}_{6c}/N&=\frac{128 m_{D^*}^2(m_{\tau}^2-m_{N_R}^2) \sqrt{\lambda_{BD^*}\lambda_{\tau N}}}{q^4} \,\re{\left(A_{12}(q^2) A_0^*(q^2)\right)} \,. \\ &\phantom{=}
		\end{aligned}
	\end{equation}
\end{widetext}
Here $N$ is the common normalization constant
\begin{equation}
	N(q^2,m_{N_R}^2)= \frac{\tau_B G_F^2 |V_{cb}|^2}{512 \pi^3 m_B^3}\sqrt{\lambda_{\tau N}}\sqrt{\lambda_{BD^*}}|C_{RR}|^2\,,
\end{equation}
and $\lambda_{BD^*}$ is abbreviation for $\lambda(\sqrt{q^2},m_B,m_{D^*})$.
Other coefficients ($I_{7,8,9}$) are 0. It follows that the differential branching fraction is 
\begin{align}
	&\frac{d\mathcal{B}(B\to D^* \tau N_R)}{dq^2}=\\
	&\frac{4 }{3 q^2} \Bigg[ 2 |A_1(q^2)|^2 (m_B + m_{D^*})^2\times \nonumber\\
	&\hspace{5em}\left(3 q^4-3 m_{\tau}^2 q^2 - 3 m_{N_R}^2 q^2 -   \lambda_{\tau N} \right) + \nonumber\\
	&\hspace{1em}\frac{64 |A_{12}(q^2)|^2 m_{D^*}^4 \left(3 q^4-3 m_{\tau}^2 q^2 - 3 m_{N_R}^2 q^2 -   \lambda_{\tau N} \right)}{q^2}\nonumber\\
	&\hspace{1em}+ \frac{2 |V(q^2)|^2 \left(3 q^4 - 3 m_{\tau}^2 q^2 - 3 m_{N_R}^2 q^2 -   \lambda_{\tau N} \right) \lambda_{BD^*}}{(m_B + m_{D^*})^2}\nonumber\\
	&\hspace{1em}\left. + \frac{3 |A_0(q^2)|^2 \left(q^4 - m_{\tau}^2 q^2 - m_{N_R}^2 q^2 -   \lambda_{\tau N} \right) \lambda_{BD^*}}{q^2} \right]\,. \nonumber
\end{align}

It is also possible to obtain expressions for the rest of the observables calculated in the section ~\ref{par:bctaunu}, namely the forward-backward asymmetry
\begin{equation}
	\begin{aligned}
		A_{\mathrm{fb}}^{D^*}=\frac{1}{\mathcal{B}(B\to D^* \tau \text{ 'inv'})}\int dq^2 (\mathcal{I}_{6c}+2 \mathcal{I}_{6s})\,,
	\end{aligned}
\end{equation}
and $\tau$ polarization asymmetry

\begin{equation}
	\begin{aligned}
		P_{\tau}^{D^*}&= \frac{1}{\mathcal{B}(B\to D^* \tau \text{ 'inv'})}\int dq^2 \frac{4  \sqrt{\lambda_{\tau N}}}{3 q^2} \times \\
		&\bigg[ 2 |A_1(q^2)|^2 (m_B + m_{D^*})^2 (2 q^2 + m_{N_R}^2 - m_{\tau}^2  )\\
		&+  \frac{64 |A_{12}(q^2)|^2 m_{D^*}^4 (-m_{\tau}^2 + m_{N_R}^2 + 2 q^2)}{q^2} \\
		&- \frac{3 |A_0(q^2)|^2 (m_{\tau}^2 - m_{N_R}^2) \lambda_{BD^*}}{q^2}\\
		&+\left. \frac{2 (-m_{\tau}^2 + m_{N_R}^2 + 2 q^2) |V(q^2)|^2 \lambda_{BD^*}}{(m_B + m_{D^*})^2} \right]\,,
	\end{aligned}
\end{equation}
which are defined in the same way as for the $B\to D$ case. Lastly, there is the longitudinal polarization fraction of $D^*$, specific to the $B\to D^*$ decay:
\begin{equation}
	\label{eq:FL_def}
	F_L^{D^*}=\frac{1}{\mathcal{B}(B\to D^* \tau \text{ 'inv'})} \;\int dq^2  \left( 2\mathcal{I}_{1c}-\mathcal{I}_{2c}\right)\,.
\end{equation}

\subsection{$B_c \to J/\psi \tau N_R$}

Note that this process is completely analogous to $B\to D^* \tau N_R$ and therefore one can use the same set of formulas for $\mathcal{B}(B_c \to J/\psi \tau N_R)$. The only difference is in the phase space (now using different masses) and in the form factors, calculated by lattice QCD~\cite{Harrison:2020gvo}.




\end{document}